\documentclass{article}
\usepackage{spconf,amsmath,graphicx}
\usepackage{amsfonts}
\usepackage{amssymb}
\usepackage{bm}
\usepackage{relsize}
\usepackage{algorithmic}
\usepackage{algorithm}
\usepackage{array}
\usepackage{subfigure}
\usepackage{textcomp}
\usepackage{stfloats}
\usepackage{url}
\usepackage{verbatim}
\usepackage{cite}
\usepackage{color}
\usepackage{makecell}
\usepackage{multirow}
\usepackage{caption}
\usepackage{enumitem}

\everydisplay={
	\abovedisplayskip=.3pt
	\belowdisplayskip=.3pt
	\abovedisplayshortskip=.3pt
	\belowdisplayshortskip=.3pt
}

\setlist{nolistsep,leftmargin=1em}

\title{Variational Bayesian Inference Based Progressive Multi-Target MIMO Sensing with Nuisance Parameters}

\name{Jiayi Yao, Shuowen Zhang}
\address{Department of Electrical and Electronic Engineering, The Hong Kong Polytechnic University\\
E-mails: jiayi.yao@connect.polyu.hk, shuowen.zhang@polyu.edu.hk}

\begin{document}
\ninept

\maketitle

\begin{abstract}
\vspace{-1mm}
This paper proposes a \emph{progressive Bayesian} multiple-input multiple-output (MIMO) sensing framework for multiple targets over multiple stages. We consider a practical yet challenging scenario where the targets' angles are unknown and random parameters to be estimated, while the targets' reflection coefficients are unknown \emph{nuisance} parameters. With an initial prior probability density function (PDF) for the targets' angles, we progressively update the prior PDF for each sensing stage as the \emph{posterior PDF} obtained from the previous stage, based on which Bayesian transmit beamforming optimization is performed to minimize the sum posterior Cram\'er-Rao bound (PCRB) in estimating the targets' angles and Bayesian sensing is performed with the help of new observations in this stage. To analytically characterize the intractable and high-dimensional posterior PDF with low complexity, we propose a \emph{variational Bayesian inference} based approach which derives a surrogate posterior PDF in \emph{closed form} with only \emph{polynomial complexity} over the number of targets, in sharp contrast to existing numerical calculation approaches with \emph{exponential complexity}. Numerical results validate the efficacy of our proposed framework in progressively refining sensing performance.\looseness=-1
\end{abstract}
\vspace{-1mm}
\begin{keywords}
Multiple-input multiple-output (MIMO) radar, progressive sensing, conditional posterior Cram\'er-Rao bound (PCRB), variational Bayesian inference.
\end{keywords}

\vspace{-5mm}
\section{Introduction}
\vspace{-3mm}
Multiple-input multiple-output (MIMO) radar is a promising technology due to its capability of flexible transmit beamforming and high-resolution target sensing \cite{li2008mimo}. Existing studies on the transmit beamforming optimization for MIMO radar generally considered the case where ``predictions'' of the unknown parameters are available, based on which the Cram\'er-Rao bound (CRB) was characterized as a function of such predictions and adopted as the optimization performance metric \cite{li2008mimo}. However, such predictions may be inaccurate or unavailable. Recently, there has been a series of works on Bayesian beamforming which leveraged the complete statistical prior distribution information of the unknown parameters. The \emph{posterior Cram\'er-Rao bound (PCRB)} was adopted as the beamforming optimization performance metric which is a lower bound of the mean-squared error (MSE) with any estimator independent of the unknown parameters' true values \cite{ghaddar2026active,liu2026precoding,ghaddar2025active,hou2026sequential,xu2023mimo1,yao2024optimal,wang2025hybrid,zheng2025beyond}. When such prior distribution information is available, Bayesian sensing can also be performed 
\cite{tipping2001sparse,donoho2009message,tzikas2008variational} via minimum MSE (MMSE), maximum \emph{a posteriori (MAP)}, etc. 
    
Existing studies typically focused on one sensing stage with fixed prior. If multiple sensing stages are available, \emph{progressive sensing}, also called active sensing \cite{sohrabi2021deep,sohrabi2022active}, can apply different transmit beamforming and sensing strategies over multiple stages based on progressively updated prior information and observations, where the posterior probability density function (PDF) obtained in the previous stage can serve as the prior PDF for the current stage. However, a systematic study on such a progressive Bayesian multi-target sensing framework is currently still lacking, which motivates our study in this paper. In particular, although progressive sensing provides opportunities for successive performance refinements, it also brings a fundamental new challenge. Note that both Bayesian beamforming and Bayesian sensing are critically dependent on the posterior PDF from the previous stage, which is notoriously difficult to characterize especially for multi-target sensing. This thus motivated numerical calculation approaches by approximating the continuous ranges of the unknown parameters via discrete grids, for which the complexity is \emph{exponential} over the number of targets and prohibitive with a large number of targets. To reduce such complexity, a variational line spectral estimation (VALSE) algorithm \cite{badiu2017variational} was proposed to approximate the complex joint posterior PDF of all unknown parameters with a tractable surrogate PDF. This idea was also extended to multi-dimensional \cite{zhang2020multidimensional} and group-sparse line spectra models \cite{moderl2024variational}. However, existing VALSE-based works focused on single-input multiple-output (SIMO) or illuminators of opportunity systems with single-dimensional transmit signals without beamforming capability, while both targets' angles and reflection coefficients are jointly estimated under specified prior distributions. For general MIMO radar systems with possible nuisance parameters (e.g., reflection coefficients), there still lacks a low-complexity approach for tractable posterior PDF characterization.

    \begin{figure}[t]
		\centering
		\setlength{\abovecaptionskip}{-0cm}
		\includegraphics[width=8.5cm]{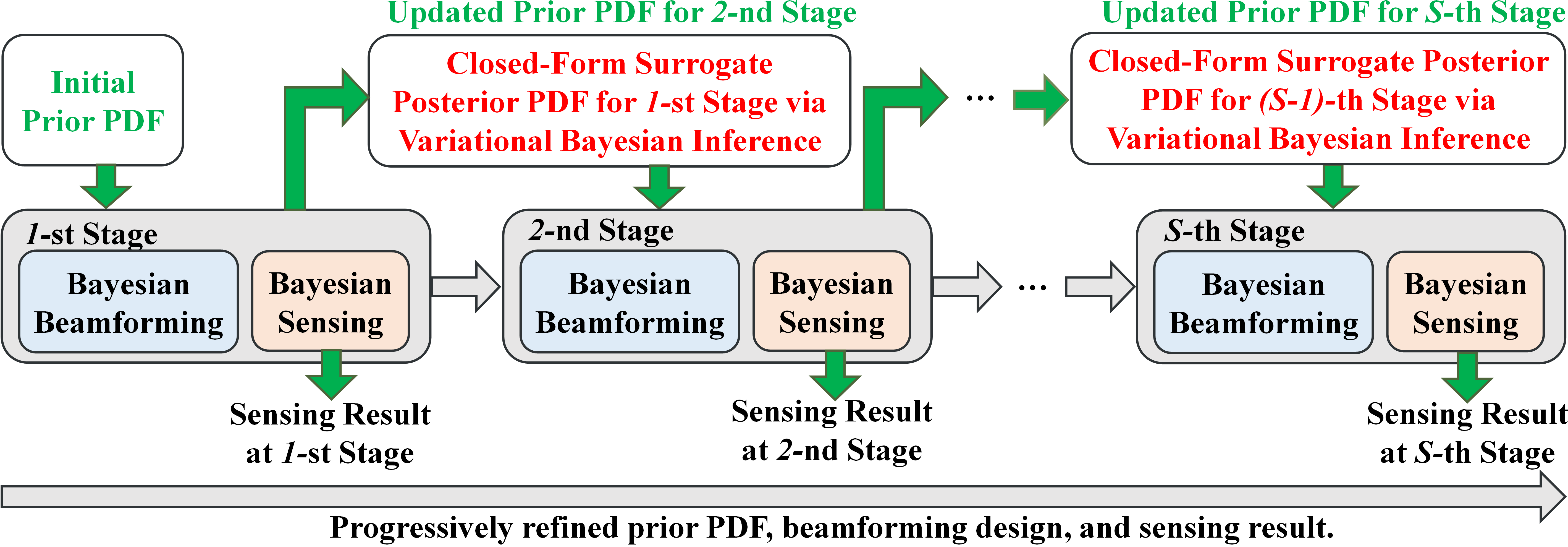}
        \vspace{-0mm}
		\caption{Progressive Bayesian multi-target sensing framework.}\label{Fig_System_Model}
        \vspace{-1mm}
	\end{figure}
This paper addresses the above challenges by establishing a scalable \emph{progressive Bayesian} multi-target MIMO sensing framework with a \emph{variational Bayesian inference} based posterior PDF characterization approach, as shown in Fig. \ref{Fig_System_Model}. We consider a practical scenario where the targets' angles are the only desired parameters, while the reflection coefficients are nuisance parameters. In each sensing stage, a multi-antenna base station (BS) sends beamformed probing signals, which are reflected by the targets and received back at the BS receiver as observations. Next, a tractable \emph{closed-form} surrogate posterior PDF is obtained via the proposed variational Bayesian inference based approach with only \emph{polynomial complexity} over the number of targets, which is then employed as the prior PDF for Bayesian beamforming optimization and Bayesian sensing in the subsequent stage. Numerical results demonstrate the progressive refinement of sensing performance via our proposed framework, and the superiority of it over various benchmark schemes.

\vspace{-7mm}
\section{System Model}\label{section_sys_model}
\vspace{-4mm}
Consider a mono-static MIMO radar system with a BS equipped with $N_t\geq1$ transmit antennas and $N_r\geq1$ co-located receive antennas under a uniform linear array (ULA) layout. The system aims to progressively sense the \emph{unknown} and \emph{random} (azimuth) angles of $M\geq1$ point targets that remain static over $S\geq 1$ stages based on the received echoes of the signals sent from the BS and reflected by the targets (i.e., \emph{observations}) as well as the prior PDF information (i.e., \emph{prior}). Let $\theta_m\in [-\frac{\pi}{2},\frac{\pi}{2})$ denote the angle of the $m$-th target and $\bm{\theta}=[\theta_1,...,\theta_M]^T$ denote the collection of all $M$ angles to be sensed. To draw fundamental insights, we consider a line-of-sight (LoS) channel between the BS and each target. Let $\alpha_{m}\!\!=\!\!\frac{\lambda}{(4\pi)^{3/2}r_m^2}e^{\!-j\frac{4\pi r_m}{\lambda}}\psi_{m}\!=\!\alpha_{m}^\mathrm{R}\!+\!j\alpha_{m}^\mathrm{I}\!\in\! \mathbb{C}$ denote the overall complex reflection coefficient for the $m$-th target which is unknown and remains unchanged over the $S$ stages, with $\lambda$ being the wavelength in meter (m), $r_m$ being the distance between the BS and the $m$-th target in m, and $\psi_{m}\in\mathbb{C}$ being the radar cross-section (RCS) coefficient. Note that $\alpha_m$'s are \emph{unknown nuisance parameters} whose collection is denoted by $\bm{\alpha}=[\alpha_1,...,\alpha_M]^T$. 
\vspace{-5.5mm}
\subsection{Observation Model}
\vspace{-3mm}
Let $\bm{a}(\theta_m)\!\!\in\!\! \mathbb{C}^{N_t\!\times\! 1}$ and $\bm{b}(\theta_m)\!\!\in\!\! \mathbb{C}^{N_r\!\times\! 1}$ denote the transmit/receive array steering vectors with $a_i(\theta_m)= e^{-\!j\pi ( N_t\!-\!2i+1 )\!\sin\! \theta _m/2},\forall i$ and $b_i(\theta_m)=e^{-\!j\pi ( N_r\!-\!2j+1 )\!\sin\! \theta _m/2},\forall i$. Let $L$ denote the number of samples in each stage, and $\bm{X}_{s}\in\mathbb{C}^{N_t\times L}$ denote the collection of probing signals sent from the transmitter in each $s$-th stage. By further assuming all $M$ targets are within the same range bin \cite{li2007range}, the collection of echo signals received in each $s$-th stage is given by
\vspace{-1mm}
\begin{align}
\bm{Y}_s=\sum_{m=1}^M\alpha_{m}\bm{b}(\theta_m) \bm{a}^H(\theta_m)\bm{X}_s+\bm{N}_s,\quad \forall s\in\mathcal{S},	\label{Y}
\end{align}
where $\bm{N}_s\in \mathbb{C}^{N_r\times L}$ with $[\bm{N}_s]_{i,j}\sim\mathcal{CN}(0,\sigma^2),\forall i,j$ denotes the circularly symmetric complex Gaussian (CSCG) noise at the BS receiver, with $\sigma^2$ denoting the average receiver noise power at each antenna, and $\mathcal{S}=\{1,...,S\}$. Let $\bm{R}_{X,s}=\frac{1}{L}\bm{X}_{s}\bm{X}_{s}^H$ denote the transmit sample covariance matrix at the $s$-th stage, which is under a per-stage transmit power constraint $\mathrm{tr}(\bm{R}_{X,s})\leq P,\ \forall s\in\mathcal{S}$.  
\vspace{-5.5mm}
\subsection{Prior Model}
\vspace{-3mm}
We consider independent targets where all the unknown parameters in $\bm{\theta}$ and $\bm{\alpha}$ are independent of each other. Let $p_{\alpha_m}(\alpha_{m})$ denote the prior PDF of $\alpha_m$. The prior PDF of the nuisance parameters in $\bm{\alpha}$ is thus given by $p_{\bm{\alpha}}\!(\bm{\alpha})\!\!=\!\!\prod_{m=1}^Mp_{\alpha_m}\!(\alpha_{m})$, which is known from targets' properties and used throughout the $S$ stages \cite{liu2026precoding}. On the other hand, the initial prior PDF of $\bm{\theta}$ before the first stage is denoted by $p_{\bm{\Theta},1}\!(\bm{\theta})$, which is assumed to be known from historic data or target appearance pattern \cite{yao2024optimal,ghaddar2026active,liu2026precoding,hou2026sequential,ghaddar2025active,xu2023mimo1,wang2025hybrid,zheng2025beyond}. Let $\bm{Y}^{(s)}\overset{\Delta}{=}\{\bm{Y}_1,...,\bm{Y}_s\}$ denote the collection of observed echo signals after the $s$-th stage. The posterior PDF of $\bm{\theta}$ at the end of each $s$-th stage is given by 
 \begin{align}\label{posterior}
p_s(\bm{\theta}|\bm{Y}^{(s)})=\frac{p(\bm{Y}^{(s)}|\bm{\theta})}{p(\bm{Y}^{(s)})}p_{\bm{\Theta},1}(\bm{\theta}),\quad \forall s\in \mathcal{S},
\end{align}
with $p(\bm{Y}^{(s)}|\bm{\theta})$ denoting the conditional PDF of $\bm{Y}^{(s)}$ given $\bm{\theta}$ and $p(\bm{Y}^{(s)})$ the PDF of $\bm{Y}^{(s)}$ which is a constant (i.e., model evidence \cite{bishop2006pattern}). Then, the prior PDF of $\bm{\theta}$ for the $(s+1)$-th stage is updated as 
\begin{align}\label{prior}
p_{\bm{\Theta},s+1}(\bm{\theta})=p_s(\bm{\theta}|\bm{Y}^{(s)}),\quad \forall s\in \mathcal{S}\backslash\{S\}.
\end{align}
Based on the updated prior PDF, transmit beamforming optimization and receiver sensing are performed. By progressively updating the prior PDF before each stage with new observations and designing the transmit beamforming accordingly, the multi-target sensing performance can be progressively refined over stages, as detailed below.

\vspace{-5mm}
\section{Proposed Progressive Bayesian Multi-Target MIMO Sensing Framework}
\vspace{-4mm}
In this section, we present our proposed progressive Bayesian multi-target MIMO sensing framework, which is illustrated in Fig. \ref{Fig_System_Model}. 
\vspace{-4mm}
\subsection{Bayesian Beamforming at the Transmitter}\label{section_PCRB_optimization} 
\vspace{-3mm}
At the start of each $s$-th stage, the BS possesses observations from the previous $s-1$ stages denoted by $\bm{Y}^{(s-1)}$ and the prior PDF of $\bm{\theta}$ denoted by $p_{\bm{\Theta},s}(\bm{\theta})=p_{s-1}(\bm{\theta}|\bm{Y}^{(s-1)})$ given by (\ref{posterior}). The BS aims to design the transmit beamforming characterized by the sample covariance matrix $\bm{R}_{X,s}$ such that the observations created in the $s$-th stage are best for sensing $\bm{\theta}$. As the MSE is difficult to be characterized explicitly, we adopt the \emph{conditional PCRB} as the beamforming optimization performance metric, which is a lower bound of the MSE with any Bayesian estimator and observation history \cite{zuo2011conditional}.

Denote $\bm{\eta}\!\!=\!\![\theta_{1}\!,...,\!\theta_{M}\!,\alpha_{1}^\mathrm{R}\!,\alpha_{1}^\mathrm{I}\!,...,\!\alpha_{M}^\mathrm{R}\!,\alpha_{M}^\mathrm{I}\!]^T$ as the real-equivalent collection of all unknown parameters. The prior PDF of $\bm{\eta}$ before the $s$-th stage is thus given by $p_{\mathrm{H},s}(\bm{\eta})\!\!=\!\!\prod_{m=1}^M{\!p_{\Theta_m,s}(\theta_m)p_{\alpha_m}(\alpha_{m})}$, where $p_{\Theta_m,s}(\theta_m)=\int p_{\bm{\Theta},s}(\bm{\theta})d\bm{\theta}_{\sim m}$ is the marginal prior PDF of $\theta_m$ and $\bm{\theta}_{\sim m}$ denotes all parameters in $\bm{\theta}$ except for $\theta_m$. The conditional PCRB for estimating $\theta_m$ at the $s$-th stage is given by \cite{zuo2011conditional}
\begin{align}
&\mathrm{PCRB}_{\theta_m,s}^{\mathrm{C}}\label{PCRB_theta_m}\\
&\!=\!\mathbb{E}_{\bm{Y}_s,\bm{\eta}}\!\left[\frac{\partial\ln p(\bm{\eta},\bm{Y}_s|\bm{Y}^{(s-1)})}{\partial\bm{\eta}}\!\left(\frac{\partial\ln p(\bm{\eta},\bm{Y}_s|\bm{Y}^{(s-1)})}{\partial\bm{\eta}}\right)^H\right]_{m,m}^{-1}. \nonumber
\end{align}
We further consider a mild and practical condition of $\mathbb{E}[\bm{\alpha}]\!=\!0$, which holds under the Swerling -0/-I/-II models for $\psi_m$ when $\psi_m$ and $r_m$ are uncorrelated \cite{swerling1960probability}. Define $\bm{M}(\theta_m)\overset{\Delta}{=}\bm{b}( \theta _m) \bm{a}^H(\theta_m)$ and let $\dot{\bm{M}}(\theta_m)$ denote its derivative. We further denote $\bm{A}_{s,m}=\int{\!\dot{\bm{M}}^H\!(\theta_m)\dot{\bm{M}}(\theta_m)p_{\bm{\Theta},s}(\bm{\theta})d\bm{\theta}}\!\succeq\! \bm{0}$, $\beta_m\!=\!\frac{2L\mathbb{E}_{\alpha_m}[|\alpha_m|^2]}{\sigma^2}$, and $\bm{F}_{\mathrm{P},s}=\mathbb{E}_{\bm{\eta}}\Big[\frac{\partial \ln ( p_{\mathrm{H},s}(\bm{\eta}))}{\partial \bm{\eta}}\big( \frac{\partial \ln (p_{\mathrm{H},s}(\bm{\eta}))}{\partial \bm{\eta}}\big)^H\Big]$. The conditional PCRB for the MSE in sensing $\theta_m$ can be then expressed as $\mathrm{PCRB}_{\theta_m,s}^{\mathrm{C}}=\frac{1}{\beta_m\mathrm{tr}(\bm{A}_{s,m}\bm{R}_{X,s})+[ \bm{F}_{\mathrm{P},s}]_{m,m}}$. Based on this, we formulate the following Bayesian transmit beamforming optimization problem to minimize the sum PCRB among all targets:
\vspace{-1mm}\begin{equation}
        \mbox{(P1)}  \min_{\bm{R}_{X,s}\succeq \mathbf{0}:\, \mathrm{tr}(\bm{R}_{X,s}) \leq P}\ \sum_{m=1}^M\!\frac{1}{\beta_m\mathrm{tr}(\!\bm{A}_{s,m}\bm{R}_{X,s}\!)\!\!+\!\![ \bm{F}_{\mathrm{P},s}]_{m,m}}.\label{P1obj}
\vspace{-1mm}\end{equation}
Note that (P1) is a convex problem and can be solved via the interior-point method \cite{boyd2004convex}. With the progressive prior PDF update, the transmit beamforming will progressively concentrate more power around the actual values of $\bm{\theta}$'s, thereby enabling better sensing accuracy.

\vspace{-5mm}
\subsection{Bayesian Sensing at the Receiver}
\vspace{-3mm}
At the end of each $s$-th stage, the BS possesses observations from the previous $s$ stages denoted by $\bm{Y}^{(s)}$ and the prior PDF of $\bm{\theta}$ denoted by $p_{\bm{\Theta},s}(\bm{\theta})$. The posterior PDF $p_s(\bm{\theta}|\bm{Y}^{(s)})$ can be then calculated from (\ref{posterior}), based on which the MMSE estimator can be obtained as $\hat{\bm{\theta}}_\mathrm{MMSE}=\int\bm{\theta}p_s(\bm{\theta}|\bm{Y}^{(s)})d\bm{\theta}$ and the marginal MAP estimator can be obtained as $\hat{\bm{\theta}}_\mathrm{mMAP}=\mathrm{arg}{\max}_{\bm{\theta}}\ p_s(\bm{\theta}|\bm{Y}^{(s)})$ \cite{doucet2002marginal,liu2013variational}.\looseness=-1 
\vspace{-5mm}
\subsection{Complexity with Numerical Posterior PDF Calculation}
\vspace{-2mm}
Note that the posterior PDF $p_{s-1}(\bm{\theta}|\bm{Y}^{(s-1)})$ or $p_{s}(\bm{\theta}|\bm{Y}^{(s)})$
needs to be known for both the calculation of $\bm{A}_{s,m}$'s and $[\bm{F}_{\mathrm{P},s}]_{m,m}$'s in Bayesian beamforming and the calculation of $\hat{\bm{\theta}}_\mathrm{MMSE}$ or $\hat{\bm{\theta}}_\mathrm{mMAP}$ in Bayesian sensing. However, the posterior PDF is a complex function of the previous observations and the prior PDF, which is intractable to be characterized analytically and explicitly. Hence, numerical calculation is generally needed by approximating the continuous range of $\theta_m$ with $\xi_\theta$ discrete grids and the continuous ranges of $\alpha_m^\mathrm{R}$ and $\alpha_m^\mathrm{I}$ with $\xi_\alpha$ discrete grids. Using this numerical posterior calculation approach, the complexities for Bayesian beamforming and Bayesian sensing are summarized in Table \ref{Table_complexity}, where all complexities increase \emph{exponentially} with the number of targets, $M$, which is prohibitive and unaffordable with a large number of targets. To cross this hurdle, we propose a novel \emph{variational Bayesian inference} based posterior characterization approach which yields an explicit analytical expression of the posterior PDF with \emph{polynomial complexity}.

\begin{table*}[t]
\centering
\caption{Complexity with Numerical Posterior PDF Calculation vs. Proposed Posterior PDF Characterization}
\vspace{-3mm}
\scalebox{1}{
	\resizebox{\textwidth}{!}{
\begin{tabular}{|cc|c|c|c}
\cline{1-4}
\multicolumn{2}{|c|}{}                                                 & Numerical Calculation & Proposed Characterization &  \\ \cline{1-4}
\multicolumn{2}{|c|}{Bayesian Beamforming}                             & $\mathcal{O}(M(\xi_\theta^M\!\!+\!\!\xi_\theta (N_t^2\!\!+\!\!1))+(2N_t^2+M)^{3.5})$                     & $\mathcal{O}(M\xi_\theta (N_t^2+1)+(2N_t^2+M)^{3.5})$   &  \\ \cline{1-4}
\multicolumn{1}{|c|}{\multirow{2}{*}{Bayesian Sensing}} & MMSE         & $\mathcal{O}(\xi_\theta^M(\xi_\alpha^{2M}sN_tN_r(L+M)+M))$    & $\mathcal{O}(I_\mathrm{iter}\mathrm{card}(\mathcal{W}_s)(I_\alpha M^3+N_tN_r^2LM^2+2\xi_\theta N_tN_rM))$    &  \\ \cline{2-4}
\multicolumn{1}{|c|}{}                                  & Marginal MAP & $\mathcal{O}(\xi_\theta^M\xi_\alpha^{2M}sN_tN_r(L+M))$                     & $\mathcal{O}(I_\mathrm{iter}[\mathrm{card}(\mathcal{W}_s)(I_\alpha M^3+N_tN_r^2LM^2+\xi_\theta N_tN_rM)+I_{\bm{\vartheta}}MN_tN_r])$    &  \\ \cline{1-4}
\end{tabular}}}\label{Table_complexity}
\vspace{-7mm}
\end{table*}

\vspace{-3mm}
\section{Proposed Variational Bayesian Inference Based Posterior PDF Characterization}\label{section_VBI}
\vspace{-3mm}
\subsection{Equivalent Transformation and Tractable Approximation}
\vspace{-3mm}
Firstly, we define \emph{pseudo angles} $\bm{\vartheta}=[\vartheta_1,...,\vartheta_M]^T$ with $\vartheta_m\overset{\Delta}{=}\frac{\pi}{2}\sin(\theta_m)$, which has a one-to-one mapping to $\theta_m$ since $\theta_m\in[-\frac{\pi}{2},\frac{\pi}{2})$ \cite{stoica2005spectral}. This yields $a_i(\vartheta_m)= e^{-\!j( N_t\!-\!2i+1 )\vartheta_m}$ and $b_i(\vartheta_m)=e^{-\!j( N_r\!-\!2j+1 )\vartheta_m}$ in the array steering vectors, thereby equivalently simplifying the observation model. The prior PDF of $\vartheta_m$ for each $s$-th stage is given by $\tilde{p}_{\Phi_m,s}(\vartheta_m)\!\!=\!\!p_{\Theta_m,s}(\mathrm{arcsin}(\frac{2\vartheta_m}{\pi}))\frac{2}{\sqrt{\pi^2-4\vartheta_m^2}}$. In the following, we focus on the posterior PDF with respect to $\bm{\vartheta}$. 

Secondly, note that the intractability of the posterior PDF mainly lies in the intractable conditional PDF $p(\bm{Y}^{(s)}|\bm{\vartheta})$. By noting that in each $s$-th stage, the prior PDFs for all previous $s$ stages are already available under our framework, we introduce an \emph{observation window} for each stage to enable a flexible trade-off between the complexity in obtaining the conditional PDF and sensing performance. Denote the observation window at the $s$-th stage as $\mathcal{W}_s=\{\rho(s),\rho(s)+1,...,s\}$ with $\rho(s)$ being the starting index of $\mathcal{W}_s$, and $\bm{Y}_{\mathcal{W}_s}=\{\bm{Y}_{\rho(s)},\bm{Y}_{\rho(s)+1},...,\bm{Y}_{s}\}$ as the collection of observations within this window. The posterior PDF can be then expressed as\footnote{Since we assume the targets' angles are posteriorly independent, (\ref{posterior2}) is an approximation of (\ref{posterior}) \cite{broderick2013streaming}. The approximation is exact when $M=1$.}
\begin{align}\label{posterior2}
p_s(\bm{\vartheta}|\bm{Y}^{(s)})\approx\frac{p(\bm{Y}_{\mathcal{W}_s}|\bm{\vartheta})}{p(\bm{Y}_{\mathcal{W}_s})}\prod_{m=1}^M\tilde{p}_{\Phi_m,\rho(s)}(\vartheta_m),\quad \forall s\in \mathcal{S}.
\end{align}
For brevity, the conditioning on $\bm{Y}^{(\rho(s)-1)}$ is omitted in all windowed conditional PDFs hereafter.

Next, we develop a \emph{variational Bayesian inference} based approach to obtain an explicit and tractable \emph{surrogate posterior PDF} denoted by $q_s(\bm{\vartheta}|\bm{Y}^{(s)})$ that approximates the posterior PDF $p_s(\bm{\vartheta}|\bm{Y}^{(s)})$ with minimum Kullback-Leibler (KL) divergence to $p_s(\bm{\vartheta}|\bm{Y}^{(s)})$. This is equivalent to maximizing the \emph{evidence lower bound (ELBO)} of $q_s(\bm{\vartheta}|\bm{Y}^{(s)})$ given by \cite{bishop2006pattern}
\begin{align}
\mathcal{LB}(q_s)\triangleq&\mathbb{E}_{q_s}\left[\ln p(\bm{Y}^{(s)},\bm{\vartheta})-\ln q_s(\bm{\vartheta}|\bm{Y}^{(s)})\right]\\
=&\mathbb{E}_{q_s}\!\left[ \ln \frac{p(\bm{Y}_{\mathcal{W}_s}|\bm{\vartheta})\prod_{m=1}^M\tilde{p}_{\Phi_m,\rho(s)}(\vartheta_m)}{q_s(\bm{\vartheta}|\bm{Y}^{(s)})}\right]+\mathrm{const}.\nonumber
\end{align}
Since directly obtaining $q_{s}(\!\bm{\vartheta}|\bm{Y}^{(s)}\!)$ that maximizes $\mathcal{LB}(q_s)$ is not viable, we apply the \emph{mean field approximation} \cite{tzikas2008variational,badiu2017variational} to factorize
\begin{align}\label{mean_field_approximation}
q_{s}(\bm{\vartheta}|\bm{Y}^{(s)})\approx\prod_{m=1}^Mq_{s,m}(\vartheta_m|\bm{Y}^{(s)}),
\end{align}
\vspace{-2mm}% 
\par\noindent% 
where $q_{s,m}(\vartheta_m|\bm{Y}^{(s)})$ is the surrogate posterior PDF of $\vartheta_m$ in the $s$-th stage. Then, by iteratively optimizing $q_{s,m}(\vartheta_m|\bm{Y}^{(s)})$ to maximize $\mathcal{LB}(q_s)$ with the surrogate PDFs for other angles fixed at each time, we can transform the original high-dimensional optimization problem on $q_{s}(\bm{\vartheta}|\bm{Y}^{(s)})$ into a series of parallel low-dimensional problems each on $q_{s,m}(\vartheta_m|\bm{Y}^{(s)})$, thereby reducing the complexity.\looseness=-1
\vspace{-5mm}
\subsection{Optimizing Surrogate Posterior PDF for Each Angle}
\vspace{-2mm}
With the surrogate posterior PDFs for other angles being fixed, we obtain $q_{s,m}(\!\vartheta_m|\bm{Y}^{(s)}\!)$ that maximizes $\mathcal{LB}(q_s)$ as follows. Define $W_{s,m}(\vartheta_m)\triangleq\mathbb{E}_{q_{s,\sim m}}[\ln\,(p(\bm{Y}_{\mathcal{W}_s}|\bm{\vartheta}) \tilde{p}_{\bm{\varTheta},\rho(s)}(\bm{\vartheta}))]$ and $\bar{p}_{s,m}(\vartheta_m)\triangleq e^{W_{s,m}(\vartheta_m)}/\int e^{W_{s,m}(\vartheta_m)}d\vartheta_m$ with $\mathbb{E}_{q_{s,\sim m}}[\cdot]$ being the expectation taken over $\prod_{i=1,i\ne m}^M{\!q_{s,i}(\!\vartheta_i|\bm{Y}^{(s)}\!)}$. The ELBO is given as $\mathcal{LB}(q_s)=-\mathcal{D}_{\mathrm{KL}}(q_{s,m}\|\bar{p}_{s,m})+\mathrm{const}$, where $\mathcal{D}_{\mathrm{KL}}(\cdot\|\cdot)$ is the KL divergence. As $\mathcal{D}_{\mathrm{KL}}(q_{s,m}\|\bar{p}_{s,m})$ vanishes if and only if $q_{s,m}( \vartheta_m|\bm{Y}^{(s)})=\bar{p}_{s,m}(\vartheta_m)$, $\mathcal{LB}(q_s)$ can be maximized by setting
    \begin{align}\label{ln_q}
        &\ln{q_{s,m}( \vartheta_m|\bm{Y}^{(s)})}\nonumber\\
        =&\mathbb{E}_{q_{s,\sim m}}[\ln{(p(\bm{Y}_{\mathcal{W}_s}|\bm{\vartheta}))}]+\ln{\tilde{p}_{\Phi_m,\rho(s)}(\vartheta_m)}+\mathrm{const}.
    \end{align} 
    Since $\ln \tilde{p}_{\Phi_m,\rho(s)}\!(\!\vartheta_m\!)$ is known \emph{a priori}, our remaining task is to find a tractable $\mathbb{E}_{q_{s,\sim m}}\![\ln(p(\!\bm{Y}_{\mathcal{W}_s}|\bm{\vartheta}))]$ for a feasible $q_{s,m}(\vartheta_m|\bm{Y}^{(s)})$.

    To deal with the conditional PDF $p(\bm{Y}_{\mathcal{W}_s}|\bm{\vartheta})$, we introduce an auxiliary PDF $r_s(\bm{\alpha}|\bm{\vartheta})$. Based on Jensen's inequality, we have
    \begin{align}\label{auxiliary_PDF}
        \ln{p(\bm{Y}_{\mathcal{W}_s}|\bm{\vartheta})}\geq \mathbb{E}_{r_s}[\ln{(p(\bm{Y}_{\mathcal{W}_s}|\bm{\eta})p_{\bm{\alpha}}(\bm{\alpha}))}-\ln{r_s(\bm{\alpha}|\bm{\vartheta})}],
    \end{align}
    where the equality holds if and only if $r_s(\bm{\alpha}|\bm{\vartheta})=p(\bm{\alpha}|\bm{Y}_{\mathcal{W}_s},\bm{\vartheta})$. Define $J_s(\bm{\vartheta},\bm{\alpha})\overset{\Delta}{=}\ln{(p(\bm{Y}_{\mathcal{W}_s}|\bm{\eta})p_{\bm{\alpha}}(\bm{\alpha}))}$, which yields $r_s(\bm{\alpha}|\bm{\vartheta})\propto e^{J_s(\bm{\vartheta},\bm{\alpha})}$. Denote $\bm{D}_t=\bm{X}_t^T\otimes \bm{I}_{N_r}$, $\bm{c}(\vartheta_m)=\bm{a}^*(\mathrm{arcsin}\frac{2\vartheta_m}{\pi})\otimes\bm{b}(\mathrm{arcsin}\frac{2\vartheta_m}{\pi})$, and $\bm{E}_t(\bm{\vartheta})=[\bm{D}_t\bm{c}(\vartheta_1),...,\bm{D}_t\bm{c}(\vartheta_M)]$. We have \looseness=-3
    \begin{align}\label{F(theta,alpha)}
        J_s(\bm{\vartheta},\!\bm{\alpha})\!\!=\!\!-\!\!\!\!\sum_{t\in\mathcal{W}_s}\!\!\!\frac{1}{\sigma^2}\!\|\mathrm{vec}(\bm{Y}_t)\!-\!\bm{E}_t(\bm{\vartheta})\bm{\alpha}\|^2\!+\!\ln{\!p_{\bm{\alpha}}(\bm{\alpha})}\!+\!\mathrm{const}.\!\!\!\!
    \end{align}
    Since directly constructing $r_s(\bm{\alpha}|\bm{\vartheta})$ for each possible $\bm{\vartheta}$ is computationally demanding, we average $J_s(\bm{\vartheta},\bm{\alpha})$ over the current surrogate PDF of $\bm{\vartheta}$, yielding $J_s(\bm{\alpha})=\mathbb{E}_{q_s^\prime}[J_s(\bm{\vartheta},\bm{\alpha})]$ and $r_s(\bm{\alpha})\propto e^{J_s(\bm{\alpha})}$, where $q_s^\prime(\bm{\vartheta}|\bm{Y}^{(s)})$ is initialized as the prior PDF at the $\rho(s)$-th stage and set as the surrogate posterior PDF from the previous iteration otherwise. Note that the expectation in $J_s(\bm{\alpha})$ only needs \emph{one-dimensional integration} over each $\vartheta_m$'s since we apply the mean field approximation in (\ref{mean_field_approximation}), which drastically reduces complexity. In the following, we replace $r_s(\bm{\alpha}|\bm{\vartheta})$ with $r_s(\bm{\alpha})$. Let
    \begin{align}\label{alpha_hat_H}
        \bar{\bm{\alpha}}_s=\mathrm{arg}{\max}_{\bm{\alpha}}\ J_s(\bm{\alpha}),\ \bm{H}_{s}=-\nabla^2_{\bm{\alpha}}J_s(\bm{\alpha})|_{\bm{\alpha}=\bar{\bm{\alpha}}_s}.
    \end{align}
    We then have $r_s(\bm{\alpha})\approx\frac{\det(\bm{H}_s)}{\pi^M}
    \exp(-(\bm{\alpha}-\bar{\bm{\alpha}}_s)^{H}\bm{H}_s(\bm{\alpha}-\bar{\bm{\alpha}}_s))$, which is exact with CSCG-distributed $\bm{\alpha}$. Denote $\bm{r}_{t,m}=\mathrm{vec}(\bm{Y}_t)-\sum_{i=1,i\ne m}^M\!\alpha_i\bm{D}_t\bm{c}(\vartheta_i),\forall t\in\mathcal{W}_s$. Based on (\ref{auxiliary_PDF}), we have
    \begin{align}\label{ln_q_s_m}
        &\mathbb{E}_{q_{s,\sim m}}\![\ln{(p(\bm{Y}_{\mathcal{W}_s}|\bm{\vartheta}))}]\!=\!\!\!\sum_{t\in\mathcal{W}_s}\!\!\!\frac{1}{\sigma^2}\!\Big[\!\!-\!\mathbb{E}_{q_{s,\sim m},r_s}[|\alpha_m|^2\|\bm{D}_t\bm{c}(\vartheta_m)\|^2]\nonumber\\[-2mm]
        &+2\mathfrak{Re}\{\mathbb{E}_{q_{s,\sim m},r_s}[\alpha_m\bm{r}_{t,m}^H\bm{D}_t\bm{c}(\vartheta_m)]\}\Big]+\mathrm{const}.
    \end{align} 
    
    Based on this, we can obtain a \emph{closed-form} $q_{s,m}(\! \vartheta_m|\bm{Y}^{(\!s\!)}\!)$ from (\ref{ln_q}) and (\ref{ln_q_s_m}). Denote $\hat{\bm{\vartheta}}^\prime_s$ as the estimate of $\bm{\vartheta}$ from the previous iteration, $\bm{\zeta}_{s,m}\!\!=\!\!\sum_{t\in\mathcal{W}_s}\!\bm{\zeta}_{s,m,t}$, and $\bm{\zeta}_{s,m,t}\!=\!\frac{2}{\sigma^2}\!(\!-\!\!\sum_{i=1,i\ne m}^M[\bm{H}_{s}^{-1}]_{i,m}\times\bm{D}_t^H\!\bm{D}_t\bm{c}(\!\hat{\vartheta}^\prime_{s,i})\!+\!\bm{D}_t^H\!\big(\!\mathrm{vec}(\bm{Y}_t\!)\!-\!\sum_{i=1,i\ne m}^M\!\bar{\alpha}_{s,i}\!\!\bm{D}_t\bm{c}(\hat{\vartheta}^\prime_{s,i}\!)\big)\bar{\alpha}_{s,m}^*)$. Define $\bm{d}(\!\vartheta_m\!)\!\!\in\!\!\mathbb{C}^{(2N_t\!+\!2N_r-3)\times 1}$ with $d_i(\vartheta_m)\!\!\overset{\Delta}{=}\!\!e^{-j(2N_t+2N_r-2-2i)\vartheta_m}$. Moreover, define $\bm{\gamma}_t^{\prime}\!\in\!\mathbb{C}^{(2N_t+2N_r-3)\times 1}$ that satisfies $\|\bm{D}_t\bm{c}(\vartheta_m)\|^2\!=\!\bm{\gamma}_t^{\prime H}\bm{d}(\vartheta_m)$. We then have
    \begin{align}\label{q_vartheta_y}
        &\ln{q_{s,m}( \vartheta_m|\bm{Y}^{(s)})}\nonumber\\[-3mm]
        =&\ln{\tilde{p}_{\Phi_m,\rho(s)}(\vartheta_m)}+\sum_{p=1}^{N_tN_r}|\zeta_{s,m,p}|\cos(c^{(p)}\vartheta_m-\mu_{s,m,p})\nonumber\\[-4mm]
        &+\sum_{q=1}^{2N_t+2N_r-3}|\gamma_{s,m,q}|\cos(d^{(q)}\vartheta_m-\mu^\prime_{s,m,q})+\mathrm{const},
    \end{align}
    \vspace{-3mm}% 
    \par\noindent% 
    where $\bm{\gamma}_{s,m}=\sum_{t\in\mathcal{W}_s}\bm{\gamma}_{s,m,t}$ with $\bm{\gamma}_{s,m,t}=-\frac{1}{\sigma^2}([\bm{H}_{s}^{-1}]_{m,m}+|\bar{\alpha}_{s,m}|^2)\bm{\gamma}_t^\prime$, $c^{(p)}\!\!=\!\!-(-N_t\!+\!N_r\!+\!2\lfloor\! \frac{p-1}{N_r}\!\rfloor\!-\!2((p\!-\!1) \mathrm{mod}\, N_r\!))$, $\mu_{s,m,p}\\=\!\!\mathrm{arg}(\zeta_{s,m,p})$, $d^{(q)}\!\!=\!\!-(2N_t\!\!+\!\!2N_r\!-\!2\!\!-\!\!2q)$, and $\mu^\prime_{s,m,q}\!\!=\!\!\mathrm{arg}(\gamma_{s,m,q})$.

    Finally, the pseudo angle $\hat{\vartheta}_{s,m}$ can be sensed via either Newton's method to maximize (\ref{q_vartheta_y}) in marginal MAP or integration of $\mathbb{E}_{q_{s,m}}[\vartheta_m]$ in MMSE, which is fed into the following iterations for obtaining the surrogate posterior PDF of other angles. By iteratively updating all $q_{s,m}(\vartheta_m|\bm{Y}^{(s)})$'s until convergence, the overall surrogate posterior PDF of $\bm{\vartheta}$ is obtained via (\ref{q_vartheta_y}), which is used for the Bayesian beamforming and Bayesian sensing in the $(s+1)$-th stage.
    
\vspace{-5mm}
\subsection{Complexity with Proposed Posterior PDF Characterization}
\vspace{-2mm}
The complexity of the proposed characterization approach mainly lies in the calculation of $\bar{\bm{\alpha}}_s$, $\bm{H}_{s}$, and $\bm{\vartheta}$. Let $I_{\bm{\alpha}}$, $I_{\bm{\vartheta}}$, and $I_\mathrm{iter}$ denote the maximum numbers of Newton iterations for updating $\bar{\bm{\alpha}}_s$ and each $\vartheta_{m}$, and the maximum number of outer iterations of the proposed approach, respectively. Denote $\mathrm{card}(\cdot)$ as the cardinality of a set. The complexities for Bayesian beamforming and Bayesian sensing with our proposed posterior PDF characterization are summarized in Table \ref{Table_complexity}. Note that all complexities are \emph{polynomial} in the number of targets, $M$, which is much lower compared to its \emph{exponential} counterpart with numerical calculation.
\vspace{-4mm}
\section{Numerical Results}\label{section_numerical_results}
\begin{figure}[t]
		\centering
        \vspace{-1mm}
		\includegraphics[width=0.46\textwidth]{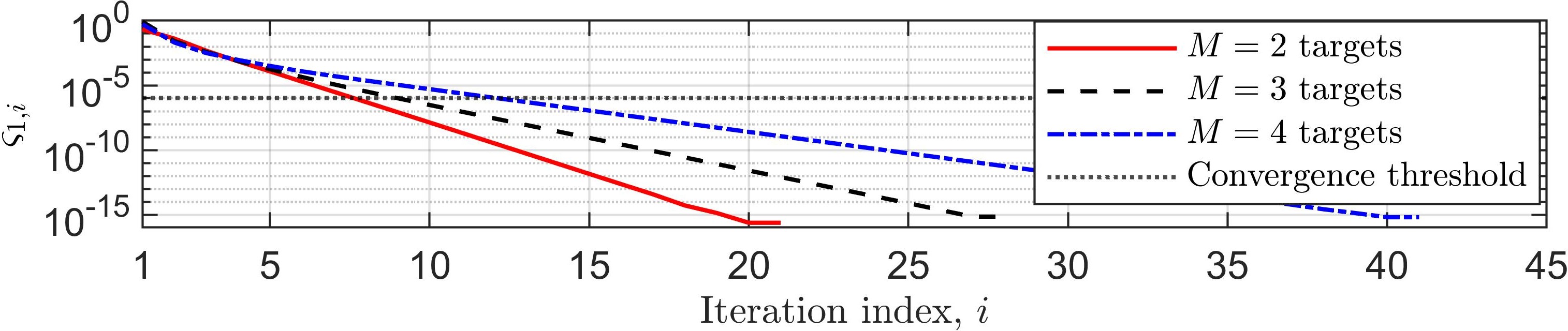}
        \vspace{-3mm}
		\caption{Convergence under $S=1$ and $PL|\alpha_m|^2\sigma^{-2}=-7$ dB.}\label{convergence_behavior}
	\end{figure}

\vspace{-3mm}
This section presents numerical results. We set $N_t=8$, $N_r=10$, $L=100$, $\sigma^2=-90$ dBm, and consider the Gaussian mixture initial prior PDF model for each $\theta_m$ as adopted in \cite{xu2023mimo1}. We further set $\alpha_m\sim\!\mathcal{CN}(0,2\times 10^{-12})$ and consider the marginal MAP estimator in our framework. Fig. \ref{convergence_behavior} shows the convergence behavior of our proposed approach in characterizing the posterior PDF under $S=1$ and different $M$, where $\varsigma_{1,i}\!\!=\!\!\|\hat{\bm{\vartheta}}^{(i)}_1\!-\!\hat{\bm{\vartheta}}^{(i\!-\!1)}_1\!\|/\|\hat{\bm{\vartheta}}^{(i\!-\!1)}_1\!\|$ denotes the relative difference of angle estimates over iterations. Fast convergence is observed in 15 iterations under convergence threshold $10^{-6}$. 

Then, we compare our proposed framework with our proposed posterior PDF characterization approach versus the following benchmark schemes, which adopt the same Bayesian beamforming but different estimation methods: 
\begin{enumerate}
\item {\bf{MUltiple SIgnal Classification (MUSIC) \cite{schmidt1986multiple}}}: The targets' angles are estimated by identifying peaks in the spatial spectrum and associated to the targets' indices according to their prior PDFs; 
\item {\bf{Estimate of Signal Parameters via Rotational Invariance Technique (ESPRIT) \cite{roy1989esprit}}}: The targets' angles are estimated via subspace-based ESPRIT, which is also used as the initialization of the first iteration in the first stage of our framework; 
\item {\bf{VALSE} \cite{badiu2017variational}}: We apply the VALSE algorithm to jointly estimate $\bm{\theta}$ and $\bm{\alpha}$ under our model where $\bm{\alpha}$ is a nuisance parameter vector; 
\item {\bf{Marginal MAP (with numerically calculated posterior PDF)}}; 
\item {\bf{MMSE (with numerically calculated posterior PDF)}}.
\end{enumerate}

    \begin{figure}[t]
		\centering
		\includegraphics[width=0.46\textwidth]{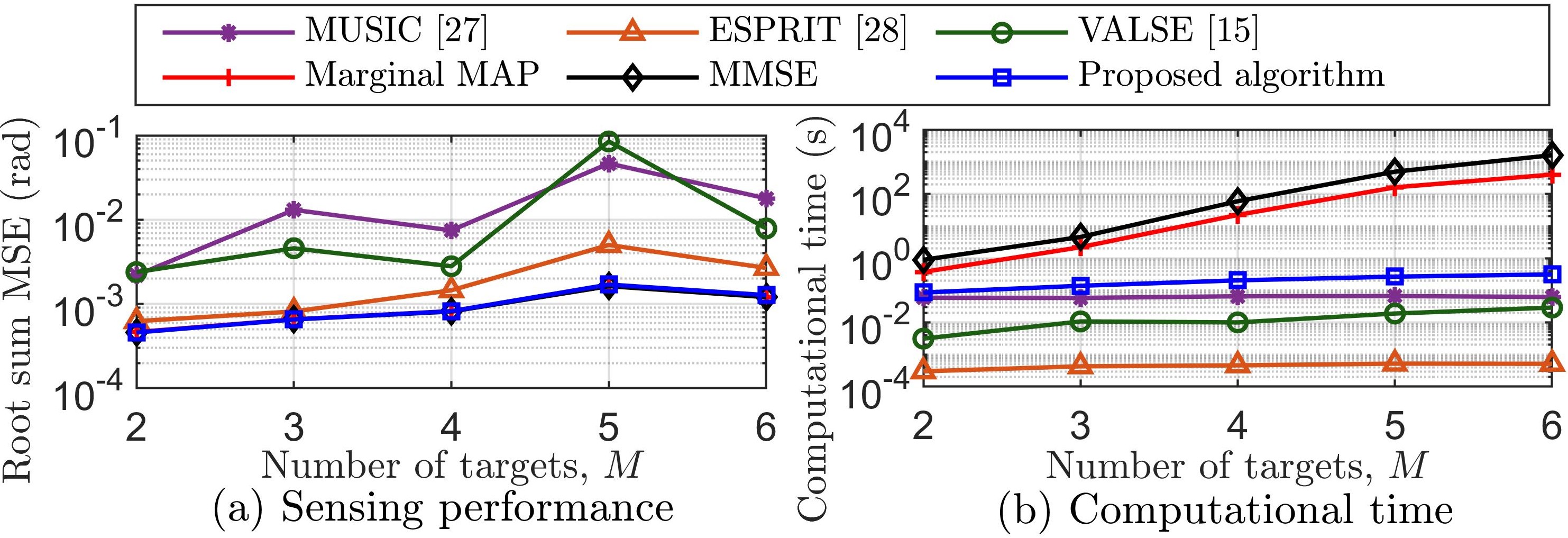}
        \vspace{-3mm}
		\caption{Root sum MSE and computational time of different schemes for $S=1$ stage and receive SNR $PL|\alpha_m|^2\sigma^{-2}=33$ dB.}\label{comp_time}
	\end{figure}

    \begin{figure}[t]
		\centering
        \includegraphics[width=0.46\textwidth]{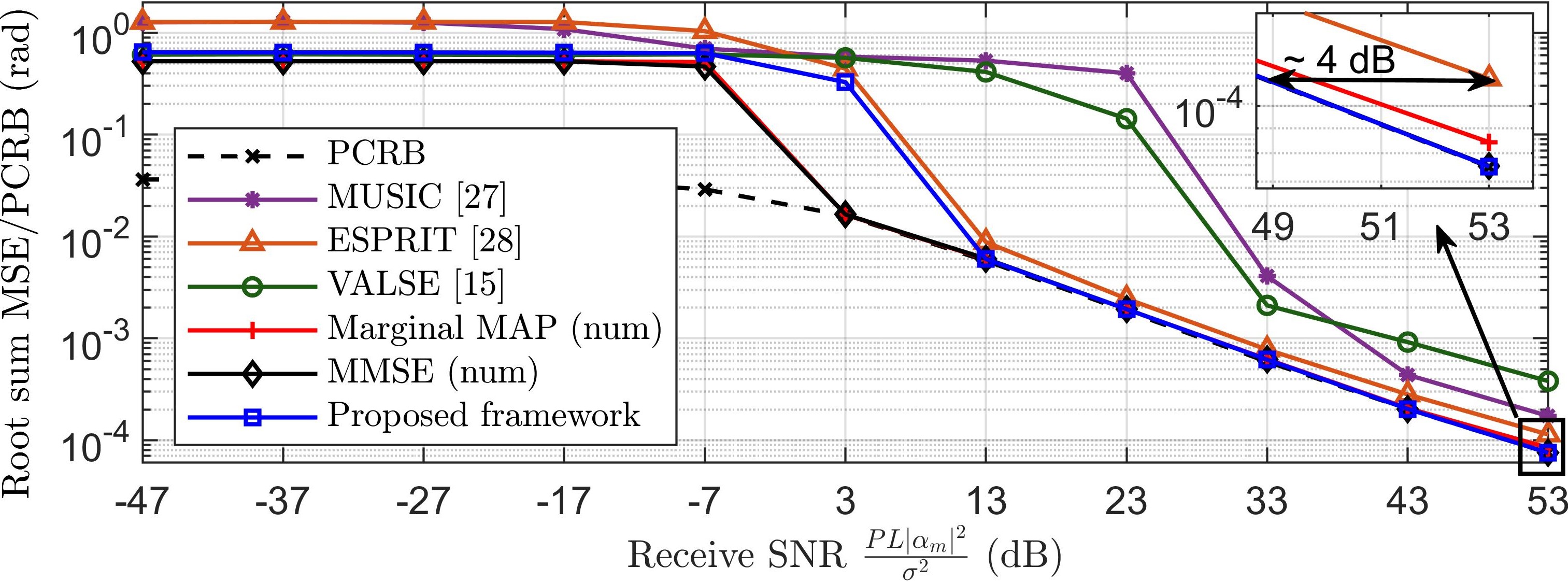}
        \vspace{-3mm}
		\caption{Root sum MSE/PCRB under $M=2$ and $S=1$.}\label{MSE}
	\end{figure}

    \begin{figure}[t]
		\centering
        \vspace{-0mm}
		\includegraphics[width=0.46\textwidth]{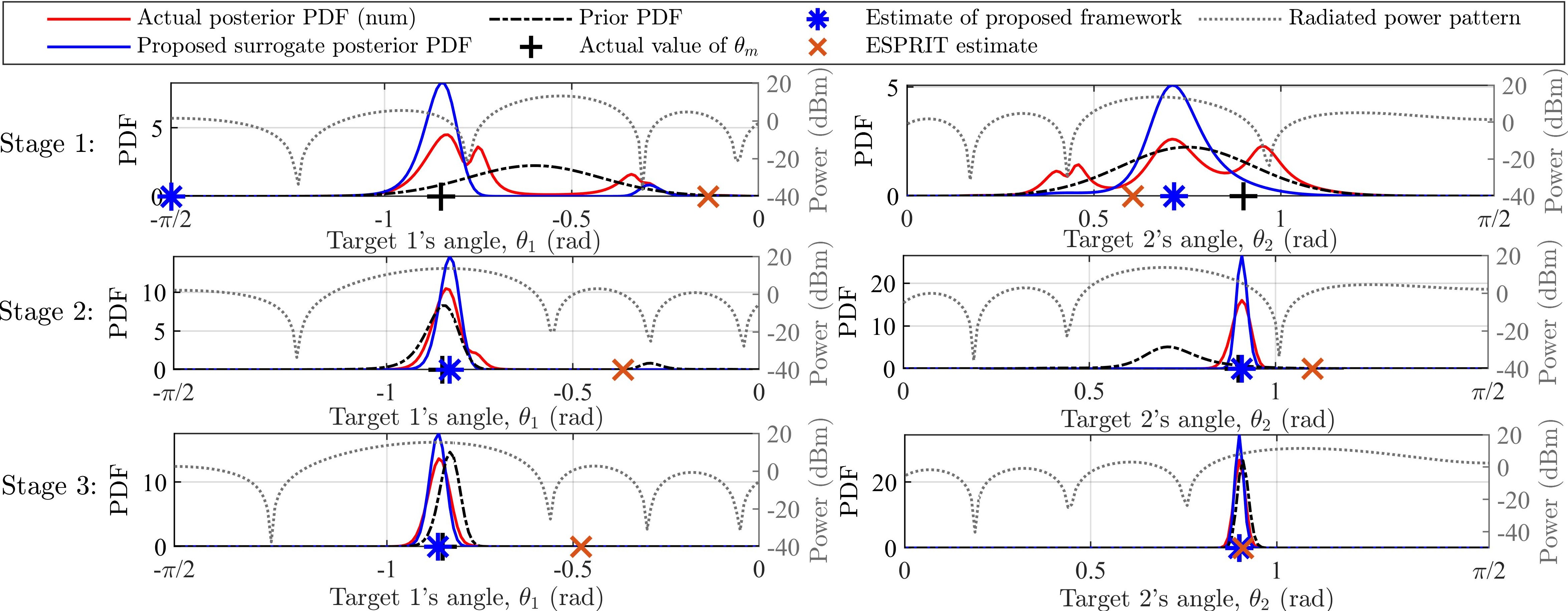}
        \vspace{-3mm}
		\caption{Prior PDFs update and sensing performance under receive SNR $PL|\alpha_m|^2\sigma^{-2}\!=\!0$ dB, $\mathcal{W}_s=\{1,...,s\}$, and $M=2$.}\label{multistage_all}
	\end{figure}

    \begin{figure}[!t]
		\centering
        \vspace{-0mm}
		\includegraphics[width=0.46\textwidth]{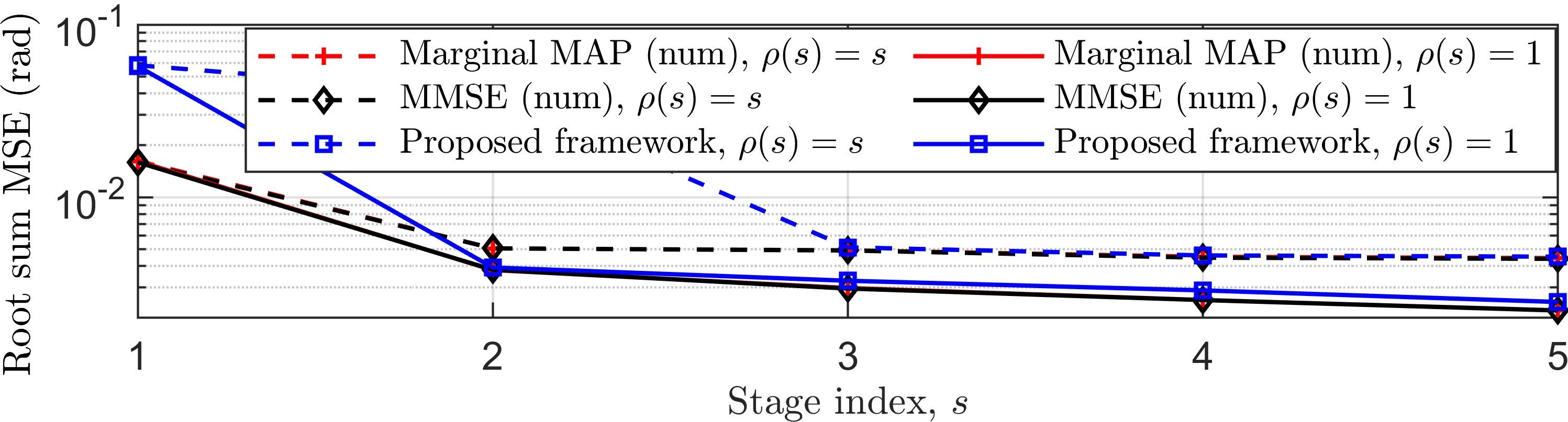}
        \vspace{-3mm}
		\caption{Sensing performance with different observation windows under receive SNR $PL|\alpha_m|^2\sigma^{-2}\!=\!13$ dB, $S\!=\!5$, and $M\!=\!2$.}\label{RMSE_stage}
	\end{figure}
 Fig. \ref{comp_time} shows the root sum MSE and the computational time of the proposed framework and the benchmark schemes versus the number of targets. It is observed that our proposed framework outperforms all schemes and achieves the same performance as that of marginal MAP and MMSE with numerical calculation of posterior PDF, while it consumes significantly less computation time compared to its numerical calculation based counterparts due to the closed-form expression of the posterior PDF, especially as $M$ becomes large. Fig. \ref{MSE} evaluates the root sum MSE of various schemes versus the receive SNR under $M=2$. It is observed that our proposed framework outperforms all schemes except for MMSE with numerically calculated posterior PDF, and approaches the latter as well as the PCRB in the moderate-to-high receive SNR regime. These results validate the effectiveness of our proposed framework. 
 
 Next, we evaluate the performance of progressive sensing under our proposed framework. Fig. \ref{multistage_all} shows the update of prior PDFs and the angle estimates. It is observed that the surrogate posterior PDFs approximate the posterior PDFs well and gradually concentrate around the actual targets' angles as the sensing stage progresses, leading to more accurate estimates. The Bayesian beamforming also tends to concentrate more power around highly-probable angles progressively. Moreover, even when the ESPRIT initialization provides inaccurate estimates, our proposed framework effectively improves them by exploiting the surrogate posterior PDFs. Finally, Fig. \ref{RMSE_stage} shows the sensing performance with different observation windows. It is observed that enlarging the observation window with smaller $\rho(s)$ improves the sensing accuracy at a cost of higher complexity.

\vspace{-5mm}\section{Conclusions}
\vspace{-3mm}
This paper proposed a variational Bayesian inference based progressive multi-target MIMO sensing framework with nuisance parameters. In each sensing stage, the BS sends beamformed probing signals designed to minimize the sum conditional PCRB for all angles based on the prior PDF, which are reflected by targets and received back at the BS as observations. Then, a variational Bayesian inference based approach was proposed to derive a closed-form surrogate posterior PDF, which is used for Bayesian sensing and Bayesian beamforming in the subsequent stage. The proposed framework only has polynomial complexity over the number of targets, which makes it highly scalable. Numerical results verified the performance and complexity advantages of our proposed framework and demonstrated the progressive refinement of sensing performance.

\newpage
\vfill\pagebreak
\bibliographystyle{IEEEtran}
\bibliography{ref}

\end{document}